\input{aipcheck}

\documentclass[
    ,final            % use final for the camera ready runs
  ]
  {aipproc}

\layoutstyle{6x9}

\begin{document}

\title{Stability of organic molecules against shocks in the young Solar nebula}

\classification{95.30.Ft, 97.82.Jw}
\keywords      {protoplanetary disk chemistry, shocks, chondrules}

\author{Inga Kamp}{
  address={Kapteyn Astronomical Institute, Groningen, The Netherlands}
}

\author{Milica Milosavljevi\'{c}}{
  address={Faculty of Mathematics, University of Belgrade, Serbia}
}

\begin{abstract}
One of the fundamental astrobiology questions is how life has formed in our Solar System. In this context the formation and stability of abiotic organic molecules such as CH$_4$, formic acid and amino acids, is important for understanding how organic material has formed and survived shocks and energetic particle impact from winds in the early Solar System. Shock waves have been suggested as a plausible scenario to create chondrules, small meteoritic components that have been completely molten by energetic events such as shocks and high velocity particle impacts. We study here the formation and destruction of certain gas-phase molecules such as methane and water during such shock events and compare the chemical timescales with the timescales for shocks arising from gravitational instabilities in a protosolar nebula.

\end{abstract}

\maketitle

%%%%%%%%%%%%%%%%%%%%%%%%%%%%%%%%%%%%%%%%%%%%
%% MAINMATTER
%%%%%%%%%%%%%%%%%%%%%%%%%%%%%%%%%%%%%%%%%%%%

\section{Introduction}

Recent observations have shown that protoplanetary disks are ubiquitous in locations where new stars are forming. At the same time, astronomers found that at least 10\% of nearby Solar like stars harbor planetary systems very different from our own, namely with Jupiter-like planets on Earth-like orbits. These findings have revived discussions on the formation of our Solar System and its uniqueness.

A fundamental question in this context is the formation of complex abiotic organic molecules in protoplanetary disks, such as CH$_4$, formic acid and amino acids. From cometary observations in our own Solar System, we know that a small fraction of the icy material consists in fact of such organic molecules. During a period of enhanced dynamic activity, the ``Late Heavy Bombardment'', evaporating and impacting comets may have delivered large quantities of water and organic material to Earth. But the young protoplanetary disks may have been a rather hostile environment for complex molecules. The protoplanetary disk is impinged by strong UV and X-ray radiation (stellar activity), and a stellar wind with highly energetic particles is driven into the disk at later stages. In addition, a massive protoplanetary disk, marginally unstable, will develop spiral waves that can lead to local shocks, in which density and temperature can be many times higher than in the unperturbed gas. Such shock waves have for example been suggested as a viable mechanism to form chondrules \citet{desch2002}. In the following, we will study the impact of a recurring shock from such a spiral wave on the gas phase chemistry.

\section{The Model}

In a typical region of planet formation, a shock arises from spiral waves that travel at a fixed pattern speed as opposed to the speed of the gas \citep{boss2005}. As a result there can be a large difference in orbital velocity between this spiral pattern and gas moving on nearly Keplerian orbits. 

We pick two points at a distance of 5~AU inside a typical T Tauri disk model \citep{kamp2004} and fix the initial temperature and density of the unperturbed gas to $T(1) = 2094$~K, $n(1) = 10^8$~cm$^{-3}$ , $\tau_{\rm UV}(1) = 2.55$ and $T(2) = 128$~K, $n(2) = 10^9$~cm$^{-3}$ , $\tau_{\rm UV}(2) = 26.5$. Within hours after the shock, temperature and density fall back to their equilibrium values. 

\subsection{Chemistry}

The time-dependent chemistry is solved using the variable-coefficient ordinary differential equation solver ``dvode'' \citep{brown1989}. The chemical network is composed of reactions from the UMIST database \citep{woodall2007} that involve 48 species \citep{kamp2004}, among which are water and CH$_4$. The chemistry has been tested against equilibrium results from \citet{kamp2004} for a protoplanetary disk as well as for a test case of a dark cloud from \citet{millar1990}.

\subsection{Shock}

%\begin{figure}
%  \includegraphics[height=.2\textheight]{shock}
%  \caption{Schematic view of the model geometry. The spiral wave rotates with its pattern speed which is locally lower than the orbital speed of the gas. This leads to the gas encountering a periodic shock every time it overtakes the spiral pattern.}
%\label{shock_sketch}
%\end{figure}

%Fig.~\ref{shock_sketch} shows a sketch of the shock in the protoplanetary disk.
\begin{figure}
  \includegraphics[height=.2\textheight]{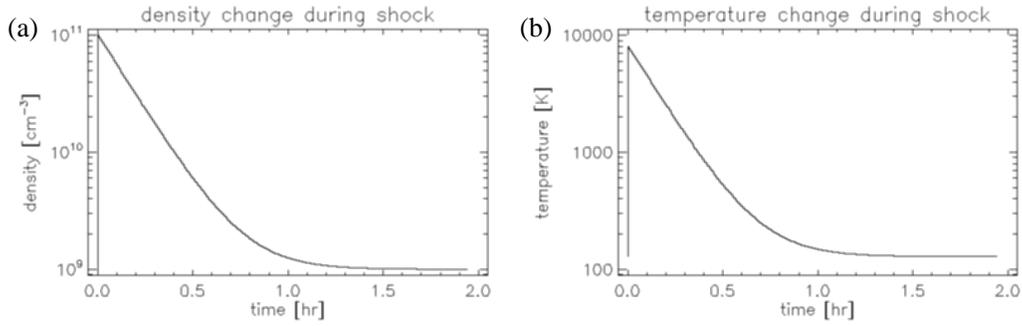}
  \caption{Density (a) and temperature (b) structure across the shock. We use the jump conditions outlined in Eq.(\ref{shock_eq1}) and (\ref{shock_eq2}) to compute the peak density and temperature of the gas and apply a fixed cooling timescale of 10 minutes.}
\label{shock_quant}
\end{figure}

We model a local gas volume element at a distance of 5~AU from the central solar-type star. The change in gas properties across the shock is defined by a set of jump conditions
\begin{eqnarray}
\label{shock_eq1}
\rho_2 & = & \rho_1 \frac{(\gamma + 1) M^2}{(\gamma -1) M^2 +2} \\
\label{shock_eq2}
T_2 & = & T_1 \frac{\left(2\gamma M^2 - (\gamma -1)\right)\left((\gamma -1)M^2 +2 \right)}{(\gamma +1)^2 M^2}
\end{eqnarray}
where $\rho_1$ and $T_1$ are the unperturbed gas density and temperature, and $\rho_2$ and $T_2$ the physical gas conditions at the shock front. $M$ is the Mach number and $\gamma$ the ratio of specific heats of the gas before and behind the shock. The latter depends on the chemical composition of the gas itself. Fig.~\ref{shock_quant} shows an example of a shock with a cooling timescale of 10~minutes. At a distance of 5~AU the relative velocity between the gas and the shock is $\sim 10$~km/s (taken from simulations of \citet{boss2005}).

\section{Timescales}

\begin{figure}
  \includegraphics[height=.22\textheight]{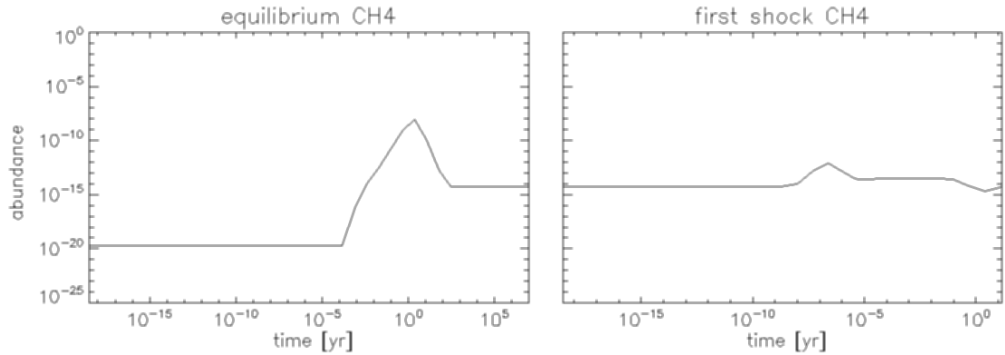}
  \caption{Time-dependent abundance of CH$_4$ for the second point in the disk model: the left panel shows the evolution until chemical equilibrium is reached, the right panel shows the change in abundance during the shock event.}
\label{CH_4}
\end{figure}

\begin{figure}
  \includegraphics[height=.22\textheight]{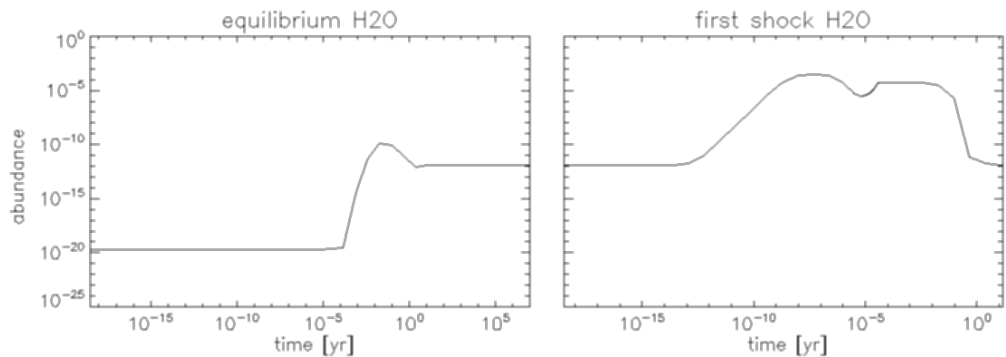}
  \caption{Time-dependent abundance of H$_2$O for the second point in the disk model: the left panel shows the evolution until chemical equilibrium is reached, the right panel shows the change in abundance during the shock event.}
\label{H2O}
\end{figure}

Our time dependent chemistry model shows that the chemistry settles back to equilibrium on timescales smaller than the shock orbital period. Fig.~\ref{CH_4} and \ref{H2O} show the time-evolution of methane and water abundance at high optical depth (second point in the disk model) towards equilibrium and during the first passage through the shock.

\section{Formation of molecules in shocks}

We discuss here only the two molecules methane and water. Both of them are formed in the shock, reaching abundances up to one thousand (methane) and a million (water) times higher than their equilibrium values under the unperturbed gas conditions.

Analysis of the models reveals that methane is predominantly formed through collisions between CH$_3$ and molecular hydrogen or water. The main destruction channels are collisions with atomic hydrogen, but also to a minor extent photodissociation into CH$_2$ and H$_2$. Water is formed via collisions between OH and molecular hydrogen, but also through electron recombination of H$_3$O$^+$. Destruction is either through collisions with atomic hydrogen (so the direct backreaction that returns the OH to the gas phase) or through ion-molecule chemistry, namely H$_2$O + Si$^+$ $\rightarrow$ SiOH$^+$ + H.

The efficient formation of water during the shocks has an important impact on the cooling timescale. This has already been noted by \citet{desch2007}. The presence of large amounts of water might lead to very fast cooling timescales of the gas behind the shock (only minutes) and this would make the condition for e.g. chondrule formation during such shock events very unfavourable. A next step would be of course the self-consistent determination of the gas temperature from a detailed heating/cooling balance of the gas.

%%%%%%%%%%%%%%%%%%%%%%%%%%%%%%%%%%%%%%%%%%%%%%%%
%% BACKMATTER
%%%%%%%%%%%%%%%%%%%%%%%%%%%%%%%%%%%%%%%%%%%%%%%%

\begin{theacknowledgments}
We would like to thank Steve Desch and Ken Rice for instructive discussions on Chondrule formation in shock events and shock physics in hydrodynamical simulations of protoplanetary disks.
\end{theacknowledgments}



\bibliographystyle{aipproc}   % if natbib is available

\bibliography{bibkamp}

\IfFileExists{\jobname.bbl}{}
 {\typeout{}
  \typeout{******************************************}
  \typeout{** Please run "bibtex \jobname" to optain}
  \typeout{** the bibliography and then re-run LaTeX}
  \typeout{** twice to fix the references!}
  \typeout{******************************************}
  \typeout{}
 }

\end{document}